\documentclass[twocolumn]{aastex701}

\usepackage{gensymb}

\begin{document}

\title{Early Results from the Coma Legacy IFU Survey (CLIFS):\\Ram Pressure Induced Shocks and Ionization in Jellyfish Tails}

\correspondingauthor{Lauren M. Foster}
\email{fostel8@mcmaster.ca}

\author[0000-0003-0214-9194]{Lauren M. Foster}
\email{fostel8@mcmaster.ca}
\affiliation{Department of Physics and Astronomy, McMaster University, 1280 Main Street West, Hamilton, ON, L8S 3L8, Canada}

\author[0000-0002-0692-0911]{Ian D. Roberts}
\email{ianr@uwaterloo.ca}
\affiliation{Waterloo Centre for Astrophysics, University of Waterloo, 200 University Avenue West, Waterloo, ON, N2L 3G1, Canada}
\affiliation{Department of Physics \& Astronomy, University of Waterloo, 200 University Avenue West, Waterloo, ON, N2L 3G1, Canada}

\author[0000-0003-4722-5744]{Laura C. Parker}
\email{lparker@mcmaster.ca}
\affiliation{Department of Physics and Astronomy, McMaster University, 1280 Main Street West, Hamilton, ON, L8S 3L8, Canada}

\author[0000-0003-4932-9379]{Timothy A. Davis}
\email{DavisT@cardiff.ac.uk}
\affiliation{Cardiff Hub for Astrophysics Research \&\ Technology, School of Physics \&\ Astronomy, Cardiff University, Cardiff, CF24 3AA, UK}

\author[0000-0003-1581-0092]{Alessandro Ignesti}
\email{alessandro.ignesti@inaf.it}
\affiliation{Astronomical Observatory of Padova, vicolo dell'Osservatorio 5, IT-35122 Padova, Italy}

\author[0000-0003-3255-3139]{Sean McGee}
\email{smcgee@star.sr.bham.ac.uk}
\affiliation{School of Physics and Astronomy, University of Birmingham, Birmingham B15 2TT, UK}

\author[0000-0001-7732-5338]{Nikki Zabel}
\email{nikki.zabel@uct.ac.za}
\affiliation{Department of Astronomy, University of Cape Town, Private Bag X3, Rondebosch 7701, South Africa}

\author[0000-0001-5880-0703]{Ming Sun}
\email{ms0071@uah.edu}
\affiliation{Department of Physics and Astronomy, University of
Alabama in Huntsville, 301 Sparkman Drive, Huntsville, AL 35899, USA}

\author[0000-0002-0587-1660]{Reinout J. van Weeren}
\email{rvweeren@strw.leidenuniv.nl}
\affiliation{Leiden Observatory, Leiden University, PO Box 9513, 2300 RA Leiden, The Netherlands}

\begin{abstract}
Jellyfish galaxies, which exhibit tails of gas opposite to their direction of motion, are a galaxy population showcasing the most extreme effects of ram pressure stripping (RPS). We present the emission line properties of a preliminary sample of five jellyfish galaxies in the Coma cluster, observed with the WEAVE Large-IFU as part of the Coma Legacy IFU Survey (CLIFS). When complete, CLIFS will form a sample of 29 jellyfish galaxies in Coma, selected based on the presence of one-sided tails in the radio continuum, enabling a comprehensive picture of the effects of ram pressure on galaxies in the Coma cluster. We extract emission line properties and confirm consistency between disk fluxes measured from WEAVE and MaNGA for galaxies with overlapping disk coverage between surveys. Comparing resolved radio and H$\alpha$-based star formation rates, we find that, in contrast to the disk, the dominant source of tail emission is not star formation. We find evidence for diffuse ionized gas excited by RPS-driven shocks in the tails, as indicated by: (1) LINER-like tail emission with the [\ion{O}{1}]/H$\alpha$ BPT diagnostic; (2) enhanced [\ion{O}{2}]/H$\alpha$ ratios in the tails relative to the disks; and (3) similarly elevated emission line velocities and velocity dispersions in the tails with respect to the disks. These results demonstrate that ram-pressure–driven shocks dominate the ionized emission in jellyfish galaxy tails.
\end{abstract}


\section{Introduction}
\label{sec:introduction}

Galaxies are a largely bimodal population in their observed properties, with either high star formation rates (SFRs), blue colours, and late-type morphologies, or low SFRs, red colours, and early-type morphologies \citep[e.g.,][]{strateva_2001,brinchmann_2004,baldry_2006,blanton_2009,peng_2010}. Galaxies transform from one population to the other by quenching (from active to passive) or more rarely, rejuvenating (vice-versa). As galaxies undergoing the transition process, known as green valley galaxies, are infrequently observed, it is assumed that this phase is relatively short-lived \citep{salim_2014}.

Galaxies in dense environments are more commonly quenched at fixed stellar mass than galaxies in the field, suggesting there are additional quenching mechanisms at work in dense environments \citep[e.g.,][]{dressler_1980,postman_1984,peng_2010}. The fraction of star-forming galaxies is suppressed by $\sim2\times$ in massive galaxy clusters \citep[e.g.,][]{wetzel_2012}.

One quenching mechanism specific to dense environments is ram pressure stripping (RPS), which can remove star-forming gas from galaxies falling into groups and clusters \citep[e.g.][]{gunn_1972,quilis_2000}. The strength of RPS scales with the density of the intra-cluster medium (ICM) and the square of the relative velocity between the galaxy and the ICM, making it most effective in galaxy clusters where the velocity dispersion and ICM density are highest. The circumgalactic medium (CGM) and outer atomic and ionized gas are stripped first, followed by a fraction of the inner molecular gas \citep{zabel_2022,brown_2023,watts_2023}. The existing stellar content remains unaffected by RPS \citep{boselli_2022}. 

An extreme class of galaxies known as jellyfish have one-sided tails of stripped gas and debris, exhibiting clear evidence of RPS. These tails are most often observed in H$\alpha$ \citep[e.g.,][]{poggianti_2017a,boselli_2018}, \ion{H}{1} \citep[e.g.,][]{kenney_2004,chung_2007}, and the radio continuum \citep[e.g.,][]{gavazzi_1987,roberts_2021}, but are detectable across the electromagnetic spectrum \citep[e.g.,][]{gavazzi_2001,sun_2006,sun_2010,sivanandam_2010,smith_2010,jachym_2014,jachym_2017}. Jellyfish galaxies offer a unique laboratory for the study of RPS, as they are the most extreme examples of galaxies affected by RPS, and the only galaxies which we know to be experiencing strong, ongoing ram pressure. In addition to gas stripping, enhanced star formation has been observed on the leading edge of jellyfish disks \citep[e.g.,][]{gavazzi_2001,troncosoiribarren_2020,lee_2016,hess_2022,zhu_2024}, where RPS compresses the interstellar medium (ISM), enabling more efficient star formation due to higher molecular gas densities \citep[e.g.,][]{schulz_2001,moretti_2020a,moretti_2020b,cramer_2021}. However, this does not seem to be a generic feature of infalling galaxies \citep[e.g.,][]{foster_2025}.

Jellyfish tails have shown signs of ongoing star formation \citep[e.g.,][]{ebeling_2014,fossati_2016,poggianti_2017a}, but many previous studies identified jellyfish based on the presence of extra-planar star-forming knots, possibly introducing a selection bias. These existing samples allow for detailed studies of the interplay between gas stripping, ICM mixing, and eventual gas cooling but do not provide robust constraints on the prevalence of star formation in RPS tails. A sample of galaxies selected without the requirement of a clearly star forming tail is needed to truly understand what fraction of jellyfish host extra-planar star formation, and what physical conditions are required for such star formation to occur. If tail emission is not entirely due to star formation, this approach can also help reveal the contribution of other physical processes such as evolved stellar populations, shocks, or ISM-ICM mixing \citep[e.g.,][]{floresfajardo_2011,zhang_2017,poetrodjojo_2018,poggianti_2019a,tomicic_2021}.

The Coma cluster provides an ideal environment for studying the effects of ram pressure, as it is the closest ($z=0.024$) massive ($M>10^{15}\text{M}_\odot$) galaxy cluster and contains an unusually high fraction of jellyfish galaxies \citep{roberts_2021}. Coma has a large halo mass, not only increasing the likelihood of galaxies becoming jellyfish, but also enabling more massive galaxies to do so. However, this also means Coma quenches galaxies efficiently, resulting in a low spiral fraction compared to other massive clusters \citep{weinzirl_2014} and therefore fewer galaxies available to become jellyfish. Ongoing sub-group merger activity in Coma \citep[e.g.,][]{fitchett_1987,bonafede_2010,brown_2011,lyskova_2019,churazov_2023} further amplifies the strength of ram pressure via enhanced ICM motions and velocity dispersion. RPS tails have been identified at a range of wavelengths in the Coma cluster, including optical and UV continuum \citep{smith_2010,roberts_2020}, narrowband \ion{N}{2} and H$\alpha$ \citep{yagi_2010}, atomic gas \citep{molnar_2022}, and the radio continuum \citep{chen_2020,roberts_2021,lal_2022}. However, a key gap in the literature is extensive integral field unit (IFU) spectroscopy of Coma galaxies undergoing RPS. \cite{roberts_2022} used MaNGA \citep{bundy_2015} spectroscopy to present evidence for enhanced star formation on the leading half of a sample of jellyfish galaxies, including 16 galaxies in the Coma cluster. However, this study focused only on the 16 of 29 known Coma jellyfish galaxies with available MaNGA data, and the limited MaNGA field of view did not capture the full extent of their stripped tails.

In this work, we present IFU observations of 5 jellyfish galaxies in the Coma cluster, 3 of which have MaNGA coverage out to $1.5r_e$. These galaxies are a subset of those identified as undergoing active RPS based on the presence of one-sided tails in LOFAR radio continuum imaging in \cite{roberts_2021}. These observations are a part of a larger survey which will observe all 29 star-forming galaxies in Coma with radio continuum tails in the LOFAR Two-metre Sky Survey \citep[LoTSS;][]{shimwell_2017,shimwell_2019,shimwell_2022}, as well as a sample of undisturbed star-forming Coma galaxies, using the William Herschel Telescope (WHT) Enhanced Area Velocity Explorer \citep[WEAVE;][]{dalton_2012,jin_2024}.

The complete sample of Coma jellyfish forms an unbiased dataset, as evidence for extra-planar star formation was not a necessary selection criteria: low frequency ($<1~\mathrm{GHz}$) radio continuum tails trace stripped cosmic ray electrons \citep[e.g.,][]{ignesti_2022,roberts_2024} which do not necessarily imply extraplanar star formation in the tail. However, this sample is likely missing jellyfish on the low-mass end, due to LoTSS sensitivity limits. It may also be missing galaxies that are more ancient infallers, as LoTSS only detects tails with ages $\lesssim100~\mathrm{Myr}$ given the synchrotron radiative lifetimes at LOFAR frequencies. With this preliminary sample, we examine SFR distributions and emission properties across disks and tails using the MaNGA Data Analysis Pipeline \citep[DAP;][]{belfiore_2019,westfall_2019}, to constrain the physical processes driving extra-planar emission. 

The outline of this paper is as follows: in Section \ref{sec:data}, we describe the IFU data and survey design. In Section \ref{sec:methods}, we describe our data reduction methods, how disk and tail regions are defined, and how SFRs are estimated, and in Section \ref{sec:results} we compare emission properties across disks and tails. In Section \ref{sec:discussion} we discuss the significance of these results, and in Section \ref{sec:summary} we summarize our findings and outline next steps for the complete survey. We adopt a flat $\Lambda$CDM cosmology with $H_0=70\text{km s}^{-1}\text{Mpc}^{-1}$, $\Omega_{M}=0.3$, and $\Omega_\Lambda=0.7$.


\section{Data}
\label{sec:data}

\subsection{The Coma Legacy IFU Survey}

\begin{figure*}
    \centering
    \includegraphics[width=\linewidth]{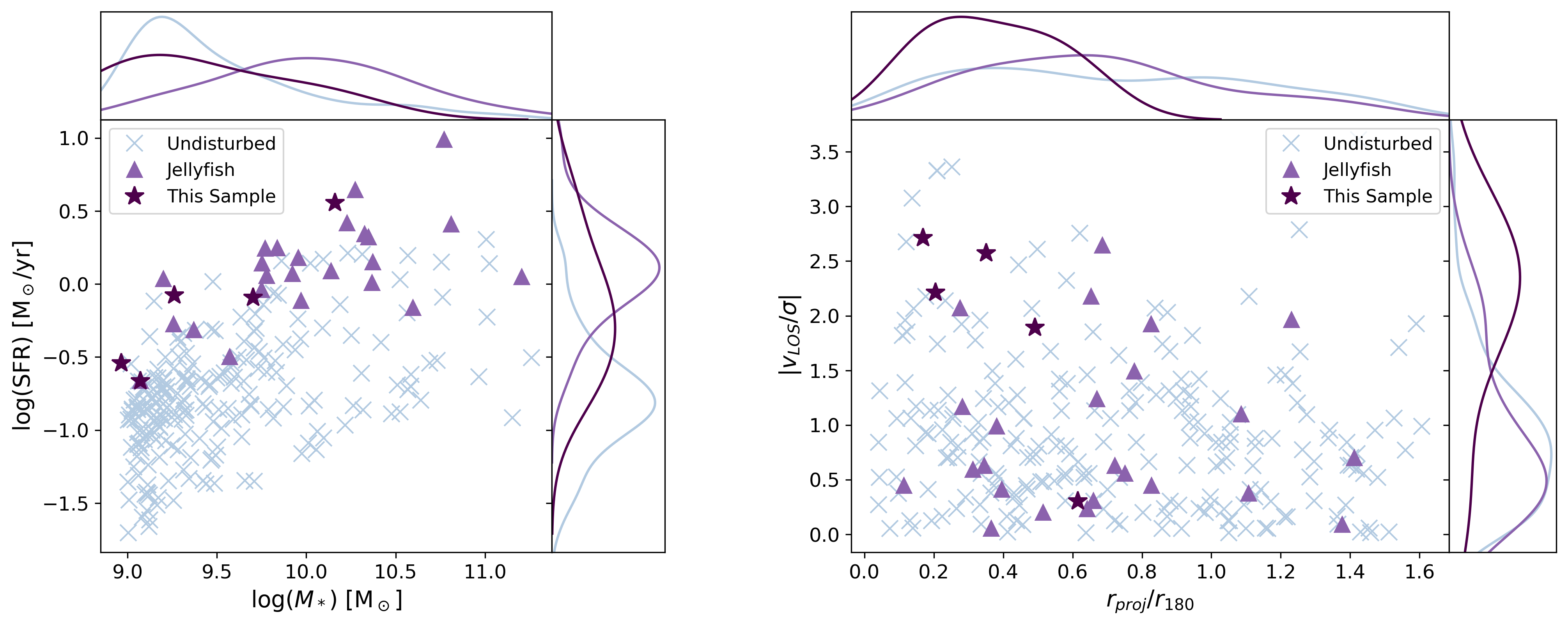}
    \caption{\textit{Left:} SFR versus stellar mass from the GALEX-SDSS-WISE Legacy catalogue \citep[GSWLC-M2;][]{salim_2016,salim_2018}, derived using UV/optical/midIR SED fitting. Apertures used likely do not contain the full extent of the RPS tails. Undisturbed Coma cluster galaxies are shown with blue Xs, LOFAR-identified Coma jellyfish are shown with light purple triangles, and the 5 jellyfish described in this paper are shown with dark purple stars. \textit{Right:} Projected phase space distribution of Coma cluster galaxies. The virial radius $r_{180}=1.85~\mathrm{Mpc}$ and velocity dispersion $\sigma=834~\mathrm{km/s}$ of the Coma cluster are calculated using equations (5) and (6) in \cite{yang_2007}, adopting a halo mass of $10^{15}\mathrm{M}_\odot$.}
    \label{figure_SFMS_PPS}
\end{figure*}

The Coma Legacy IFU Survey (CLIFS) is an ongoing survey that aims to target all 29 SDSS star-forming galaxies in the Coma cluster with radio continuum tails as identified in \cite{roberts_2021}, as well as a number of undisturbed star-forming galaxies. The complete sample covers a wide range of stellar masses and SFRs, and covers the entire spatial extent of the cluster (Figure \ref{figure_SFMS_PPS}), allowing us to probe a diverse range of galaxy types, environments, and stages of RPS. Our current sample is biased toward regions of strongest ram pressure: at high relative line-of-sight velocity near the cluster centre. Basic properties of our preliminary sample of five are provided in Table \ref{table}.

\begin{deluxetable}{ccccccccc}
\label{table}
\tablewidth{0pt}
\tablecaption{Galaxy sample coordinates and basic properties}
\tablehead{
    \colhead{CLIFS ID} & \colhead{Name} & \colhead{RA [\degree]} & \colhead{Decl. [\degree]} & \colhead{$\log M_*~[\text{M}_\odot]$ \tablenotemark{a}} & \colhead{$\log \text{SFR}~[M_\odot \text{yr}^{-1}]$\tablenotemark{a}} & \colhead{$\theta$ [\degree]\tablenotemark{b,c}} & \colhead{$i$\tablenotemark{b}} & \colhead{Hubble type\tablenotemark{d}}
}
\startdata
    39 & NGC 4858 & 194.758636 & 28.1157 & 10.160 & 0.557 & 28 & 0.81 & SBb \\
    82 & KUG1255+275 & 194.577606 & 27.3108 & 9.702 & -0.092 & 51* & 0.54* & Im \\
    151 & D100 & 195.038086 & 27.8665 & 9.262 & -0.076 & -8 & 0.71 & SBa\\
    219 & KUG1257+278 & 194.915863 & 27.5765 & 9.071 & -0.663 & -25 & 0.76 & SBm \\
    247 & LEDA83749 & 195.139786 & 27.5041 & 8.964 & -0.542 & 47 & 0.67 & Im \\
\enddata
\tablenotetext{a}{From the GSWLC-M2 \citep{salim_2016,salim_2018}}
\tablenotetext{b}{From \cite{simard_2011}, unless starred}
\tablenotetext{c}{Position angle in degrees CCW from the +x-axis}
\tablenotetext{d}{From the NASA/IPAC Extragalactic Database}
\end{deluxetable}

The WEAVE Large-IFU \citep[LIFU;][]{dalton_2012,jin_2024} is well-suited for this survey, as the large FOV of $90''\times78''$ ($\sim44~\mathrm{kpc}\times38~\mathrm{kpc}$ at Coma) covers the entire extent of each galaxy - including its tail - with a single pointing, with the exception of NGC 4848, which will require two pointings. In addition, the $2.6''$ diameter fibres and $3.4''$ centre-to-centre fibre spacing probes physical scales of 1--2~kpc at the distance of the Coma cluster ($z=0.024$). Each LIFU pointing was observed for a single one-hour block consisting of three 18-minute science exposures, using the low resolution blue+red mode (LIFU LR) which provides a spectral resolution of $R\sim2500$.

\subsection{The LOFAR Two-metre Sky Survey}

We use radio continuum imaging from the LOFAR \citep{vanhaarlem_2013} Two-metre Sky Survey \citep[LoTSS;][]{shimwell_2017,shimwell_2019,shimwell_2022} to identify tail regions as described in Section \ref{sec:methods:regions}, and to compute radio-based SFRs as described in Section \ref{sec:methods:sfrs}. LoTSS is an ongoing survey that will ultimately image the entire northern sky using the LOFAR High Band Antenna (HBA) at 120-168 MHz, with an angular resolution of $6''$ ($\sim3~\mathrm{kpc}$ at Coma) and a median noise level of $95~\mu\text{Jy/beam}$. The large primary beam size of $\sim4 \degree$ at $144~\text{MHz}$ allows for the entirety of most low-redshift clusters to be observed with a single 8-hour pointing, reaching a typical depth of $\sim100~\mu\text{Jy/beam}$. For further survey details we refer the reader to \cite{shimwell_2017} for the observing strategy, and \cite{shimwell_2019} and \cite{shimwell_2022} for the first and second data releases respectively.

\section{Methods}
\label{sec:methods}

\subsection{Datacube Reduction and Analysis}
\label{sec:cube_analysis}

We obtain Level 1 reduced data products (\texttt{CASUVERS v0.91}) from the WEAVE Operational Repository\footnote{\url{http://casu.ast.cam.ac.uk/weave/}}. For each galaxy, this includes stacked data cubes for the blue- and red-arms of the WEAVE LIFU. For both spectrograph arms, we apply the flux calibration stored in the \texttt{SENSFUNC} extension of the data files. We then downsample both the spatial and spectral axes to pixel widths of $1\arcsec$ and channel widths of $1\,\mathrm{\AA}$, with the $1\arcsec$ pixel size still oversampling the $2.6\arcsec$ fibre diameters in the LIFU. Next, we stitch together the blue- and red-arm cubes with inverse-variance weighting. This gives a single, final data cube with continuous (outside of two chip gaps) spectral coverage between $3700\,\mathrm{\AA}$ and $9400\,\mathrm{\AA}$.

The final CLIFS sample will include a mix of both WEAVE and MaNGA data cubes, which necessitates that the WEAVE and MaNGA products be analysed in a similar fashion. We use the MaNGA DAP \citep{belfiore_2019,westfall_2019} to extract stellar and emission line properties from the WEAVE datacubes; here we do not comment on the specifics of the pipeline, and instead refer the reader to \citet{belfiore_2019} and \citet{westfall_2019}. Our data closely match the MaNGA spectral and spatial resolution, allowing for direct comparison between datasets.

\begin{figure*}
    \centering
    \includegraphics[width=\linewidth]{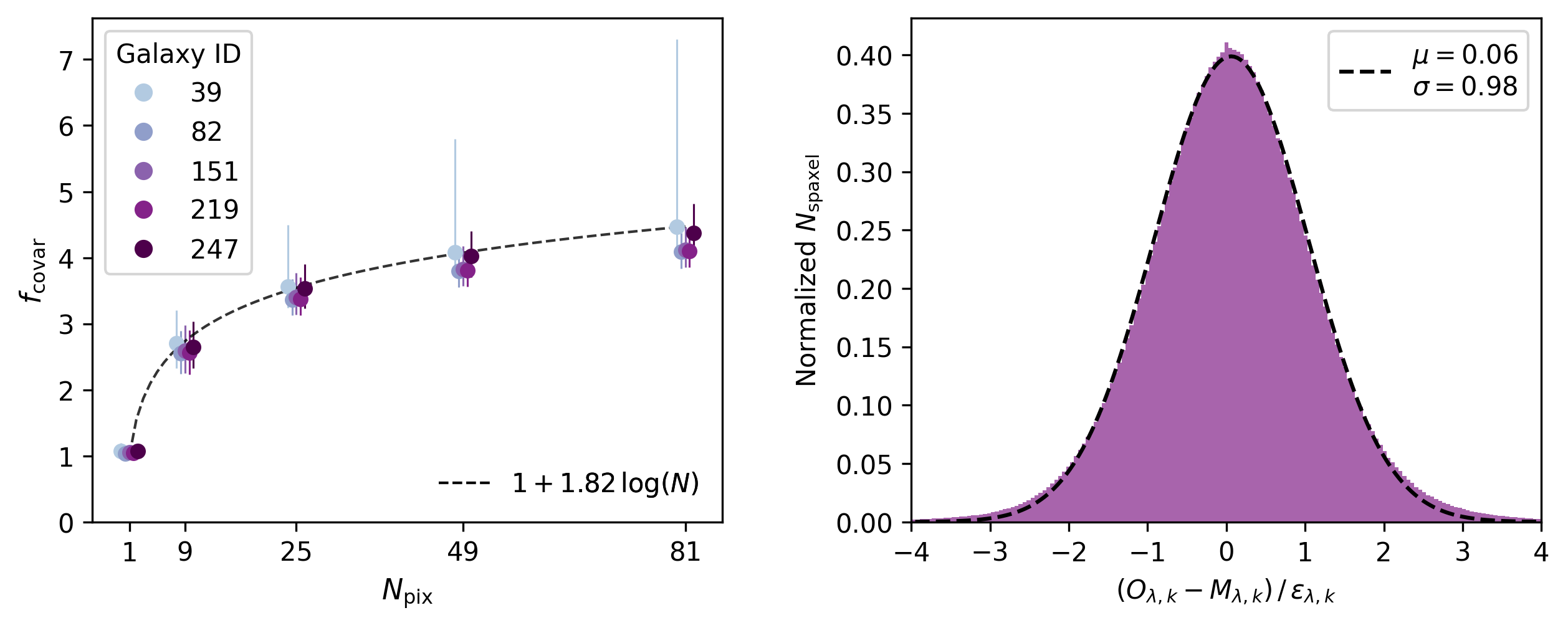}
    \caption{\textit{Left:} Covariance calibration factor as a function of bin size (see text for details). Data points correspond to median values and the error bars span between the 16th and 84th percentile, with colours distinguishing the five galaxies in this work. \textit{Right:} Residuals between the observed ($O_\lambda$) and best-fit model spectrum ($M_\lambda$), for each wavelength ($\lambda$) and spaxel ($k$) with $g$-band signal-to-noise greater than 3, with all 5 galaxies included. Residuals are normalized by the error spectrum ($\epsilon_\lambda$). The dashed line shows the best-fit Gaussian distribution.}
    \label{fig:dataquality}
\end{figure*}

We run the MaNGA DAP with the \texttt{HYB10-MASTARHC2-MASTARSSP} analysis scheme. As in MaNGA data, stellar kinematics maps are Voronoi binned to have signal-to-noise (S/N) $\sim10$ in the $g$-band, but emission line fits are performed on individual spaxels. For Voronoi binning we only include pixels that have an average per channel S/N of at least two in the $g$-band, while MaNGA uses a cut of one. However, we find that reducing the cut to a S/N of one does not affect our results. We account for pixel covariance by using the `calibrate' option in the MaNGA DAP. This method parametrizes the correction factor, $f_\mathrm{covar}$, needed to rescale noise vectors computed without accounting for covariance (see e.g.,\ \citealt{garcia_benito_2015, westfall_2019}), as

\begin{equation}
    \label{eq1}
    f_\mathrm{covar} = 1 + \alpha \log N,
\end{equation}

\noindent where $N$ is the number of pixels in a bin and $\alpha$ is a free parameter determined from the data. We measure $f_\mathrm{covar}$ by comparing the noise level directly measured from the stacked spectrum within a bin to the noise determined by propagating from the \texttt{IVAR} extension in the data cubes under the assumption of no covariance. In Figure \ref{fig:dataquality} we show an example of this calibration for our data with a best-fit $\alpha$ of ${\sim}1.8$ without strong variation from galaxy to galaxy. We refer the reader to \cite{garcia_benito_2015} and \cite{westfall_2019} for details on the procedure for fitting to equation (\ref{eq1}).

\begin{figure*}
    \centering
    \includegraphics[width=\textwidth]{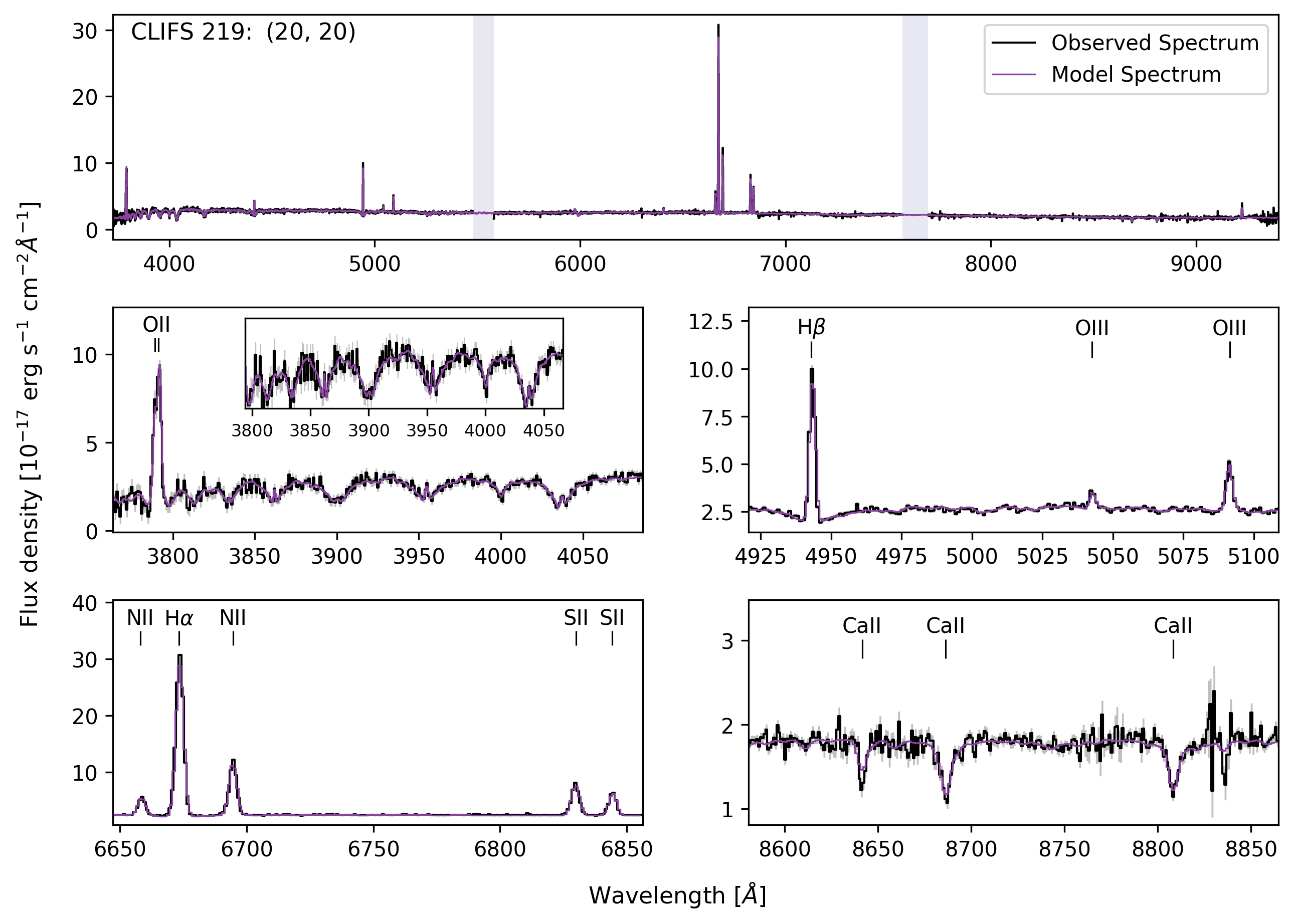}
    \caption{Example observed spectrum (black) and errors (grey regions) and best-fit model (purple) for the central spaxel of CLIFS 219. The top panel shows the full spectrum, with shaded regions marking empty areas in the spectrum from chip gaps. Key emission and absorption features are highlighted in the lower panels: [\ion{O}{2}] in the centre left, H$\beta$ and [\ion{O}{3}] in the centre right, [\ion{N}{2}], H$\alpha$, and [\ion{S}{2}] in the bottom left, and the \ion{Ca}{2} triplet in the bottom right.}
    \label{fig:specfit}
\end{figure*}

To ensure that the error spectra contained in the final science cubes are reliable, we consider the difference between the observed ($O_\lambda$) and best-fit model ($M_\lambda$) spectra, normalized by the error spectrum ($\epsilon_\lambda$). In Figure \ref{fig:dataquality} we show the resulting distribution which is well described by a best-fit Gaussian distribution centred on 0.06 with a standard deviation of 0.98, nearly identical to the expectation for random, uncorrelated noise. Finally, in Figure \ref{fig:specfit} we show an example of the MaNGA DAP fit to the WEAVE spectrum for the central spaxel of CLIFS 219 (KUG1257+278).

As the DAP is known to have difficulty modelling low signal-to-noise H$\alpha$ emission due to underlying stellar absorption \citep{belfiore_2019}, which is central to our analysis of faint tail emission, we tested several pre-processing strategies prior to running the data through the pipeline. We found that spectral smoothing did not affect the measured fluxes in either the disk or tail regions. Spatial smoothing, both with and without Nyquist re-sampling, increased the measured tail flux but led to loss of critical spatial information. Moreover, the origin of the additional flux is ambiguous, as it may reflect contamination from disk emission that was smoothed into the tail region, rather than genuine tail signal.

We also tested spectral stacking in the disk and tail regions separately, fitting two composite spectra per galaxy. This method recovers H$\alpha$ fluxes consistent within uncertainties of those obtained from the standard spaxel-by-spaxel fitting described above, and we find that variance-weighting and/or smoothing the composite spectra does not significantly increase the recovered flux. We note that while smoothing the stacked spectra improves the fit in the low signal-to-noise tail regions, it does not result in a significant change in recovered flux. As none of these efforts improved our measured fluxes, we do not pre-process the data prior to input to the DAP.

\begin{figure}
    \centering
    \includegraphics[width=\linewidth]{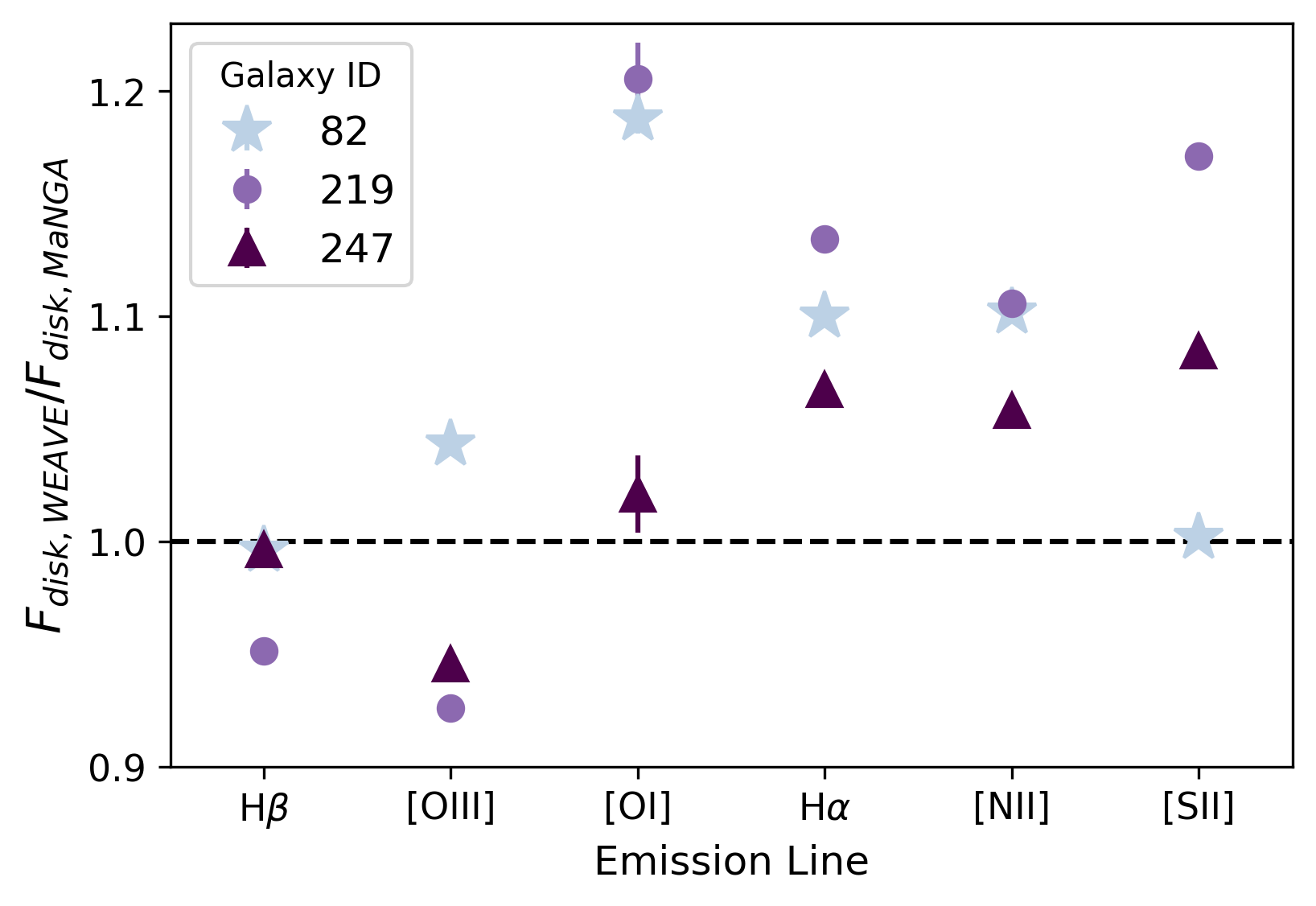}
    \caption{Ratio of disk flux measurements between WEAVE and MaNGA for BPT emission lines. Point shapes and colours distinguish the three galaxies with both CLIFS and MaNGA data. Errors on the ratios are included, but are smaller than the points in most cases. The dashed horizontal line at unity indicates where flux measurements are equivalent between the two surveys.}
    \label{figure_fluxes}
\end{figure}

16 of the 29 radio tail galaxies have existing IFU spectroscopy from the MaNGA survey; however these observations cover only the inner $\sim1.5r_e$, preventing analysis of tail emission. We compare the disk fluxes of BPT emission lines \citep[{H$\beta$ (4862\AA), [\ion{O}{3}] (5008\AA), [\ion{O}{1}] (6302\AA), H$\alpha$ (6564\AA), [\ion{N}{2}] (6585\AA), [\ion{S}{2}] (6718,6732\AA)};][]{baldwin_1981} for the 3 of 5 galaxies in the current sample with MaNGA data in Figure \ref{figure_fluxes}. The mean ratio between WEAVE and MaNGA disk fluxes is 1.1 with a scatter of 0.1. So, while the WEAVE fluxes are typically higher, the two surveys are in general agreement. We note that while the differences in [\ion{O}{1}] fluxes are higher in CLIFS 82 and 219, these discrepancies arise from the outer disk; at small radii, the flux values are in good agreement. We show SDSS $gri$ images, H$\alpha$ maps, and LOFAR radio continuum maps in Figure \ref{figure_images}.

 
\subsection{Defining Disks and Tails}
\label{sec:methods:regions}

\begin{figure*}
    \centering
    \includegraphics[width=\linewidth]{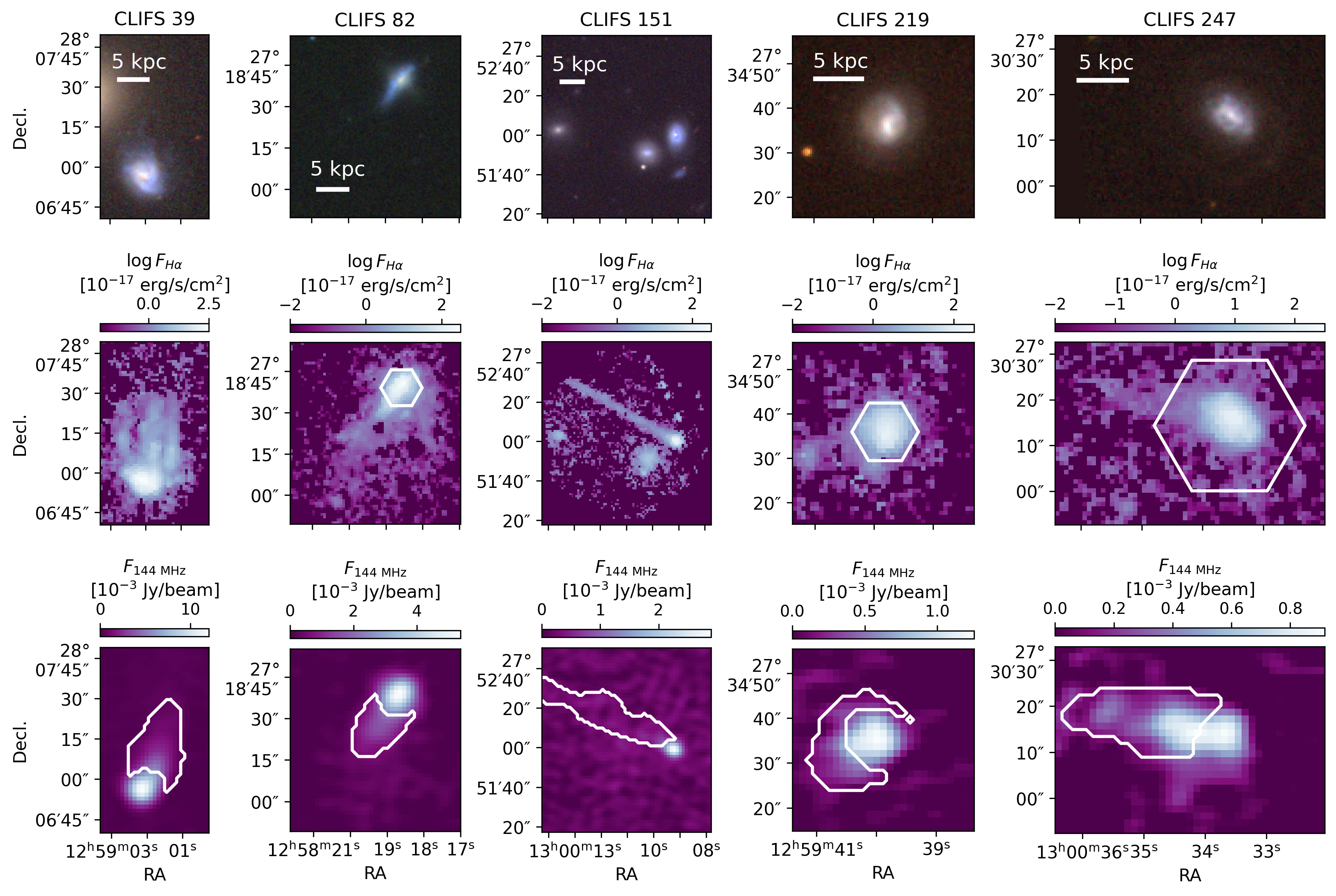}
    \caption{Top row: RGB images of each galaxy in the CLIFS sample created using SDSS \textit{g}, \textit{r}, and \textit{i}-band images and \texttt{astropy.visualization.make\textunderscore lupton\textunderscore rgb}. Middle row: WEAVE H$\alpha$ flux maps, obtained from the MaNGA DAP for the same fields of view. Overlaid are the MaNGA FOVs for the galaxies with both MaNGA and WEAVE coverage. Bottom row: LOFAR radio continuum flux maps for the same fields of view. Overlaid are the tail regions defined in Section \ref{sec:methods:regions}.}
    \label{figure_images}
\end{figure*}

We define galaxy disks using ellipses centred on the galaxy SDSS coordinates with ellipticities and position angles taken from the $r$-band Sérsic parameters provided in Table 3 of the \cite{simard_2011} morphology catalogue. We define semi-major axes of twice the effective radii in the $r$-band, chosen to encompass the full stellar disk while minimizing contamination from extraplanar emission. In the case of CLIFS 82, the \cite{simard_2011} Sérsic fit is aligned with a possible outflow feature oriented perpendicular to the main disk, evident in the SDSS image (Figure \ref{figure_images}) and confirmed by the MaNGA stellar rotation map. We instead use \texttt{imfit} \citep{erwin_2015} to fit a Sérsic profile to the SDSS $r$-band image ourselves, restricting the position angle to be approximately perpendicular to that in the \cite{simard_2011} catalogue.

As the sample was selected based on having well-defined radio tails, we use LOFAR radio continuum maps to define the tail regions. We first smooth the radio image using a Gaussian kernel with a standard deviation of 2 pixels, and mask the disk region defined above. We then use \texttt{photutils} to detect sources at $1\sigma$ above the mean in the smoothed version of the image, masking only the disk and other bright sources in the image. This threshold was chosen to fully encompass the tails while minimizing background contamination. Any region of the detected tail that overlapped with the pre-defined disk region is excluded in the final tail region. Due to the outflow-like in CLIFS 82, we instead define a more conservative elliptical tail region by eye in the LOFAR image (shown in Figure \ref{figure_images}).

\subsection{Star Formation Rates}
\label{sec:methods:sfrs}

We convert H$\alpha$ and radio continuum fluxes in the disks and tails to star formation rates (SFRs) using established empirical calibrations. Prior to computing the H$\alpha$-based SFR, we correct the H$\alpha$ flux for dust attenuation using the Balmer decrement, taking the intrinsic ratio between H$\alpha$ and H$\beta$ to be 2.86 for Case B recombination, suitable for star-forming regions ($T=10^4\text{K}$, electron density $n_e=10^4~\text{cm}^{-3}$) \citep{osterbrock_2006}. We correct each disk and tail pixel individually, as using a global value for the H$\alpha$/H$\beta$ flux ratio has been shown to underestimate the dust-corrected H$\alpha$ flux in MaNGA-like observations \citep{asari_2020}. We make a signal-to-noise cut of 3 on both lines, and if the H$\beta$ flux does not satisfy this criteria in a given spaxel, we do not dust-correct the H$\alpha$ in that spaxel. We assume:

\begin{equation}
    F_{\text{H}\alpha,\text{em}}=F_{\text{H}\alpha,\text{obs}}e^{-\tau}
\end{equation}

\noindent where $F_{\text{H}\alpha,\text{em}}$ is the emitted H$\alpha$ flux, $F_{\text{H}\alpha,\text{obs}}$ is the observed, dust attenuated H$\alpha$ flux, and $\tau$ is the optical depth at H$\alpha$, given by:

\begin{equation}
    \tau=\frac{-1}{1-q}\ln\left[\frac{F_{\text{H}\alpha,\text{obs}}/F_{\text{H}\beta,\text{obs}}}{F_{\text{H}\alpha,\text{int}}/F_{\text{H}\beta,\text{int}}}\right]
\end{equation}

\noindent where $q=1.38$ is the value of the attenuation curve at H$\beta$ \citep{calzetti_2000}, and $F_{\text{H}\alpha,\text{int}}/F_{\text{H}\beta,\text{int}}=2.86$ is the intrinsic Balmer decrement described above.

To derive the SFR from the dust-corrected H$\alpha$ emission, we adopt the \cite{kennicutt_2012} relation \citep{hao_2011,murphy_2011}:

\begin{equation}
    \text{SFR}_{\text{H}\alpha} = 5.4\times10^{-42} L_{\text{H}\alpha}
\end{equation}

\noindent where $L_{\text{H}\alpha}$ is the H$\alpha$ luminosity of the galaxy in $\text{erg}~\text{s}^{-1}$ yielding a SFR in units of $M_\odot~\text{year}^{-1}$.

To compute the radio-based SFR, we use the \cite{gurkan_2018} calibration for LOFAR $144~\text{MHz}$ imaging:

\begin{equation}
    \text{SFR}_{144~\mathrm{MHz}} = 10^{(\log_{10}(L_{144~\mathrm{MHz}})-22.06)/1.07}
\end{equation}

\noindent where $L_{144~\mathrm{MHz}}$ is the monochromatic radio luminosity of the galaxy in $\text{W}~\text{Hz}^{-1}$, also yielding a SFR in units of $M_\odot~\text{year}^{-1}$. We note that \cite{gurkan_2018} use the \cite{chabrier_2003} initial mass function (IMF), while \cite{kennicutt_2012} use the \cite{kroupa_2003} IMF; however, their SFR calibrations are effectively identical when converted to a \cite{chabrier_2003} IMF \citep{kennicutt_2012}. We also test the \cite{wang_2019} and \cite{smith_2021} calibrations, and find that they provide nearly identical results to the \cite{gurkan_2018} calibration. When computing $\text{SFR}_{144~\mathrm{MHz}}$, we mask pixels with a signal-to-noise less than 3.

\section{Results}
\label{sec:results}

\subsection{SFR Comparison}

\begin{figure}
    \centering
    \includegraphics[width=\linewidth]{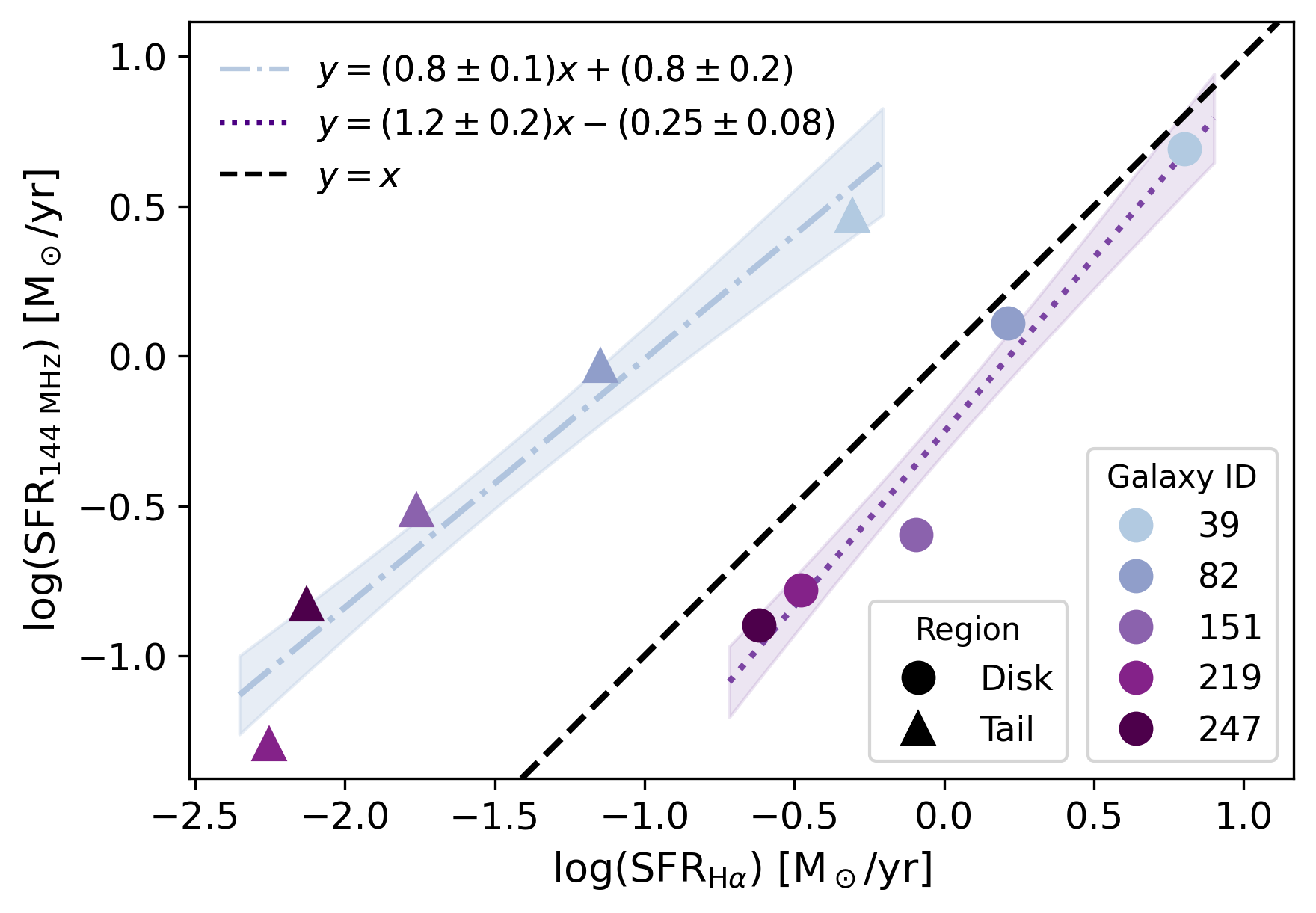}
    \caption{Radio-based SFR vs dust-corrected H$\alpha$-based SFR for the galaxies in our sample. Circles represent galaxy disks, triangles represent galaxy tails, and colours represent individual galaxies. The black dashed line shows the 1:1 relation, the purple dotted line shows the fit to the galaxy disks, and the blue dashed-dotted line shows the fit to the galaxy tails, with equations given in the legend.}
    \label{figure_SFR}
\end{figure}

We first compare galaxy SFRs derived from H$\alpha$ emission to those from the radio continuum (Figure~\ref{figure_SFR}). H$\alpha$ traces recent star formation on timescales of $\lesssim 20~\text{Myr}$ by probing ionized \ion{H}{2} regions around high-mass stars \citep{kennicutt_1998}. The $144~\text{MHz}$ radio continuum originates from synchrotron emission from cosmic ray electrons accelerated by supernovae, and thus also traces recent high-mass star formation \citep{gurkan_2018}, although the electrons can radiate for $\sim 100~\text{Myr}$ at LOFAR frequencies. However, if cosmic rays are accelerated by some other means, such as RPS, the radio-derived SFR may exceed that inferred from H$\alpha$. The same effect can emerge also if the SFR declines on a timescale shorter than the cosmic ray electron cooling time \citep{ignesti_2022,edler_2024}. The RPS scenario can also be confirmed using spectral indices: the majority of jellyfish tails show progressively steeper spectral indices as distance from the disk increases, consistent with aging cosmic ray electrons being stripped from the galaxy disk \citep{vollmer_2004,muller_2021,venturi_2022,ignesti_2023,roberts_2024b}.

We do not consider SFR uncertainties, as the intrinsic scatter of the empirical calibrations far exceeds the measurement errors on the fluxes themselves. Distinct linear fits to disk and tail data are computed using \texttt{scipy.optimize.curve\textunderscore fit}, and are given in Figure \ref{figure_SFR}. We find that disk SFRs are approximately consistent with a $1:1$ relation with an offset between indicators. In contrast, tail SFRs are systematically higher in the radio than H$\alpha$ ($\sim6-20\times$), indicating a significant contribution from some process other than star formation. However, the slope is sub-unity, meaning that as the SFR increases, the indicators reach better agreement: when the H$\alpha$ SFR is low, the radio emission and thus the radio ``SFR'' is dominated by aged cosmic ray electrons stripped from the disk, but as the H$\alpha$ SFR increases, star formation becomes a more significant contributor. These results, coupled with the spectral index results of \cite{roberts_2024b}, provide strong evidence that the synchrotron emission in RPS tails primarily originates from the galaxy disk, rather than in situ star formation.

\subsection{Characterizing Tail Emission}

\begin{figure*}
    \centering
    \includegraphics[width=\linewidth]{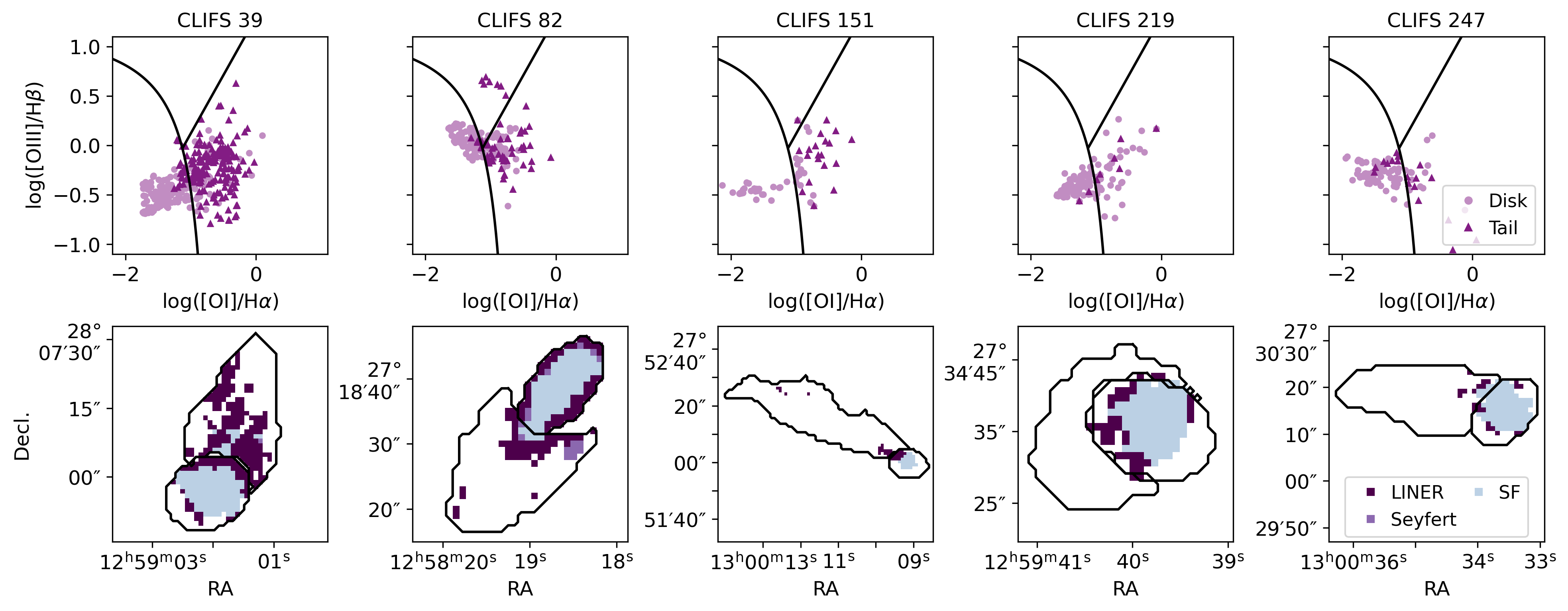}
    \caption{Top row: Resolved [\ion{O}{1}]/H$\alpha$ BPT diagram for each galaxy in the sample. Disk pixels are represented by light purple circles, and tail pixels are represented by dark purple triangles. A signal-to-noise cut of 2 has been applied to each line (H$\beta$, [\ion{O}{3}], [\ion{O}{1}], H$\alpha$), excluding pixels which do not meet this criteria in any emission line. Overlaid are the classification boundaries defined to separate AGN from star-forming sources \citep[lower LHS;][]{kewley_2001}, and to distinguish Seyfert emission from LINER emission \citep[upper and lower RHS respectively;][]{kewley_2006}. Bottom row: Galaxy maps coloured according to the BPT classification above, with x and y-axes denoting arcseconds from the SDSS centre of the galaxy, and black contours outlining the disk and tail regions. Star-forming, Seyfert, and LINER pixels are coloured blue, light purple, or dark purple respectively. Galaxy tails contain fewer pixels as the emission lines typically have lower signal-to-noise ratios.}
    \label{figure_BPT}
\end{figure*}

In Figure \ref{figure_BPT}, we show the [\ion{O}{3}]/H$\beta$ versus [\ion{O}{1}]/H$\alpha$ BPT diagrams \citep{baldwin_1981} for each galaxy, overlaid with the classification boundaries defined in \cite{kewley_2001} (AGN versus star-forming) and \cite{kewley_2006} (Seyfert versus LINER). We also show galaxy maps with spaxels coloured according to their classifications in this diagram; the equivalent figures for the [\ion{N}{2}] and [\ion{S}{2}] diagrams are given in Appendix \ref{appendix}. A signal-to-noise cut of 2 is made on the BPT lines present in the diagram (for Figure \ref{figure_BPT}: H$\beta$, [\ion{O}{3}], [\ion{O}{1}], H$\alpha$); spaxels which do not meet this criteria in any emission line are discarded, limiting the number of tail spaxels we can use for analysis in some of the galaxies in our sample. However, we note that our results are robust against the choice of signal-to-noise cut within the range 1-3.

In Figure \ref{figure_BPT}, we see that disk emission is dominated by star formation, while tail emission is dominated by LINER-like emission. The [\ion{O}{1}] BPT diagnostic is unique in this respect, with the [\ion{N}{2}] and [\ion{S}{2}] diagrams indicating that tails are either dominated by star formation or composite emission. We emphasize that here we highlight the [\ion{O}{1}] simply because the results are an anomaly compared to those from the [\ion{N}{2}] and [\ion{S}{2}] diagrams (see Appendix \ref{appendix}): the tail emission is dominated by LINER emission \textit{only} in the [\ion{O}{1}] diagnostic.

\begin{figure*}
    \centering
    \includegraphics[width=\linewidth]{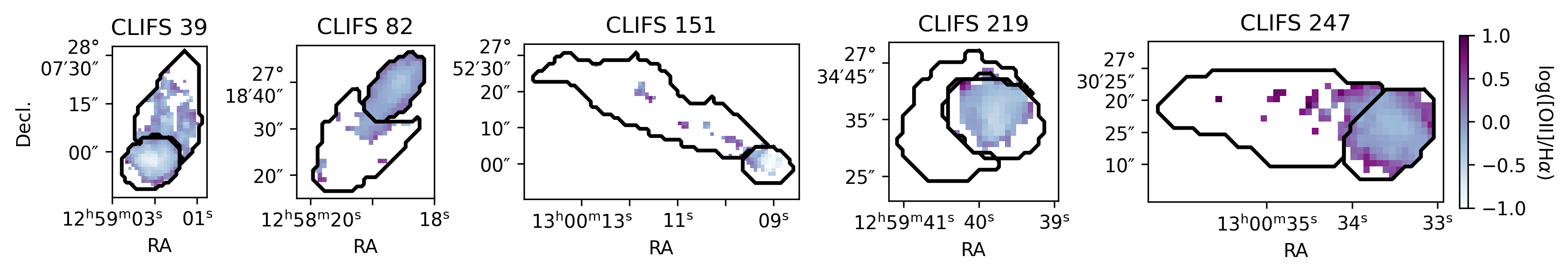}
    \caption{[\ion{O}{2}]/H$\alpha$ flux ratio maps for each galaxy in the sample, with a signal-to-noise cut of 2 applied to both lines. Black contours outline the disk and tail regions, and darker colours indicate the region is dominated by [\ion{O}{2}] emission.}
    \label{figure_oii_ha}
\end{figure*}

Following \cite{moretti_2022}, we examine logged resolved [\ion{O}{2}]/H$\alpha$ flux ratios in Figure \ref{figure_oii_ha}, making a S/N cut of 2 on each line, where negative values indicate the region is dominated by H$\alpha$ emission and hence star formation. Positive values indicate that [\ion{O}{2}] is dominant, which is atypical for normal star forming regions \citep[e.g.,][]{martin_1997}. As the [\ion{O}{2}] lines (3737~\r{A} and 3729~\r{A}) lie near the edge of our spectral range (beginning at 3660~\r{A}), the signal-to-noise in the tails is poor (average $\mathrm{S/N}\sim1.2$), resulting in limited coverage of these regions in some galaxies. However, in the galaxies with sufficient numbers of pixels with $\mathrm{S/N}>2$ (CLIFS 29, 82, 247), we see that the disks are dominated by H$\alpha$ emission, while the tails show significantly higher [\ion{O}{2}]/H$\alpha$ ratios, confirmed statistically with the two-sample Kolmogorov–Smirnov and Anderson-Darling tests. Our values in both the tails and disks are comparable to those reported by \cite{moretti_2022} for a sample of 13 jellyfish in Abell 2744 and Abell 370 using MUSE data.

\subsection{Velocity Maps}

\begin{figure*}
    \centering
    \includegraphics[width=\linewidth]{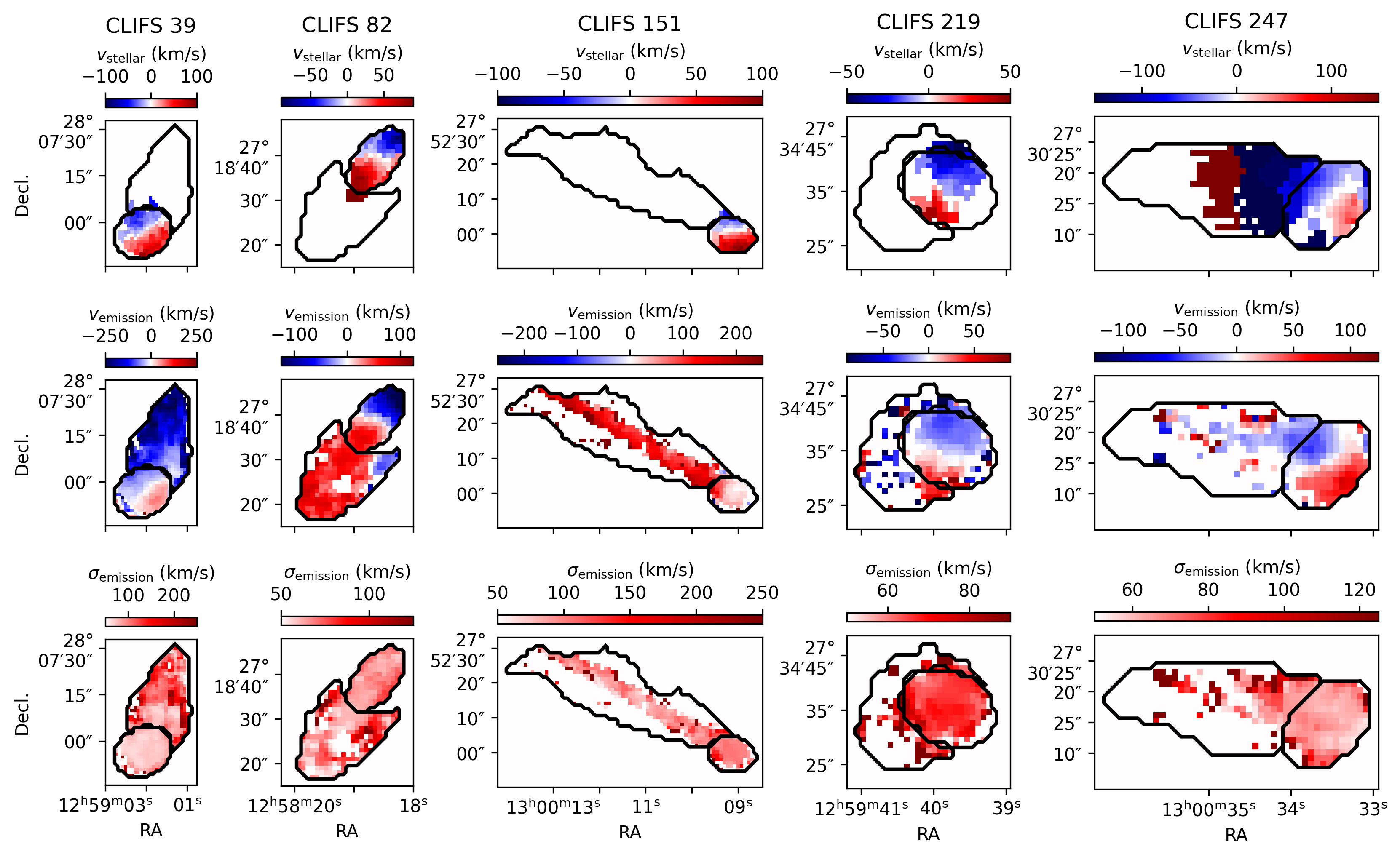}
    \caption{Stellar velocity (top row), emission line velocity (middle row), and H$\alpha$ velocity dispersion (bottom row) maps for each galaxy in the sample, with black contours outlining the galaxy disk and tail regions. Redder colours indicate higher positive velocities, bluer colours indicate higher negative velocities, with white representing net 0 velocity.}
    \label{figure_vmaps}
\end{figure*}

Finally, in Figure \ref{figure_vmaps} we show the stellar velocity, emission line velocity, and H$\alpha$ velocity dispersion maps, with a signal-to-noise cut of 2 applied to each measurement. We also apply a signal-to-noise cut of 2 on H$\alpha$ flux for the emission line velocity and emission line velocity dispersion maps. The MaNGA DAP sets all emission line velocities equal in their fitting process, resulting in a single emission line velocity map \citep[see][]{belfiore_2019}. Velocity dispersions, however, are fitted independently for each line; here we show the H$\alpha$ dispersion map, though results are consistent across all BPT lines. We note that due to the $R=2500$ spectral resolution, spaxels with velocity dispersions less than $50$ km/s have been excluded from analysis. 

In the disk regions, the magnitude of the emission-line velocities per spaxel are typically higher than stellar velocities by a factor of $\sim1.7$. RPS is expected to increase gas velocities relative to the stellar component, with the increase aligned with the stripped tail and redshifted or blueshifted depending on the galaxy's direction of motion relative to the line of sight. Distinguishing gas that has been accelerated by RPS from undisturbed disk gas requires structural analysis of each galaxy \citep[e.g.,][]{souchereau_2025}, which we defer to future work with the full CLIFS sample. 

In the tail regions, stellar velocity maps are generally empty, indicating there is no significant stellar population in this area, consistent with RPS leaving existing stellar populations largely unaffected. CLIFS 247 exhibits high stellar velocities in the tail region, however we note that this originates from the DAP fitting the Voronoi-binned continuum emission of many low-signal spaxels that are associated with faint, extended spiral arm features around the galaxy. Emission line velocities and H$\alpha$ velocity dispersions are elevated in the tails relative to the disks, especially at their edges, corresponding to the most recently stripped gas. ISM-ICM interactions can further accelerate the gas in the stripped tail by injecting turbulence down to sub-kpc scales \citep{li_2023,ignesti_2024}. We again confirm that the absolute values of the emission line velocity and H$\alpha$ velocity dispersion distributions in the tails versus disks are statistically distinct using the two-sample Kolmogorov–Smirnov and Anderson-Darling tests.

\section{Evidence for Shock-Driven Emission}
\label{sec:discussion}

Figure~\ref{figure_SFR} indicates that processes other than just star formation are taking place in the tails of Coma cluster jellyfish galaxies, as the radio SFRs exceed those derived from H$\alpha$, whereas in the disks the two SFRs agree. The diagnostics presented in Section \ref{sec:results} collectively suggest that diffuse ionized gas (DIG), likely produced by shocks driven by ram pressure, is the dominant contributor to the observed radio continuum and line emission in the tails, as discussed in the following subsections.

\subsection{Elevated [\ion{O}{1}]/H$\alpha$ Ratios}

First, we examined the spatially resolved emission line ratios using standard BPT diagnostics. In particular, the [\ion{O}{1}]/H$\alpha$ diagram indicates that tail regions are LINER-like, consistent with emission from non-stellar sources. Although the origin of LINER emission remains debated, it is commonly attributed to ionization from evolved stars, low-luminosity AGN, and/or shocks \citep[e.g.,][]{heckman_1987,filippenko_1992,dopita_1995,singh_2013}, where shocks are the most likely contributor for RPS tails. In any case, this suggests that star formation is not the dominant ionization source in the stripped tails.

Our findings are consistent with results from other jellyfish galaxy studies. \citet{poggianti_2019b} found that the BPT classification of jellyfish tails varies significantly depending on the diagnostic used. In their sample of 16 cluster galaxies with extended H$\alpha$ tails in the  GAs Stripping Phenomena \citep[GASP;][]{poggianti_2017a} survey, LINER-like emission was found to be more prominent in the [\ion{O}{1}] diagram than in the [\ion{N}{2}] or [\ion{S}{2}] diagrams. While the fraction of LINER-classified spaxels varies from galaxy to galaxy and across the classification diagram considered, 3/16 galaxies in their sample have significantly high [\ion{O}{1}]/H$\alpha$ ratios.

Building on \citet{poggianti_2019b}, \citet{tomicic_2021} analyze the BPT emission of a larger sample of 41 GASP galaxies across various stages of RPS. They introduce a metric to quantify the offset from the \cite{kewley_2001} star-forming/LINER dividing line in BPT space. They find that galaxy tails are dominated by diffuse ionized gas emission (92\% DIG), whereas galaxy disks are divided between dense and diffuse emission (60\%). They also find that that DIG regions in the tail are largely LINER dominated, whereas DIG regions in the disk are more evenly divided between star-forming and LINER. The dense gas emission in both regions is indicative of star formation, as expected.

Similar studies using MUSE have also found elevated [\ion{O}{1}]/H$\alpha$ ratios: in a case study of an extreme jellyfish \citep{fossati_2016}, an H$\alpha$/X-ray orphan cloud \citep{ge_2021}, and a sample of seven jellyfish galaxies in the Leo cluster \citep[A1367;][]{pedrini_2022}. Several other jellyfish galaxy studies use BPT diagnostic diagrams to distinguish between emission sources \citep[e.g.,][]{poggianti_2017b,peluso_2022}, although these are limited to the [\ion{N}{2}] and [\ion{S}{2}] BPT diagnostics due to spectral range constraints, or focus exclusively on jellyfish hosting central AGN, making poor comparisons for our sample.

\subsection{Elevated [\ion{O}{2}]/H$\alpha$ Ratios}

The enhanced [\ion{O}{2}]/H$\alpha$ ratios observed in the tails further support the presence emission driven by sources other than star formation. High [\ion{O}{2}]/H$\alpha$ ratios can be interpreted as arising from low gas densities \citep{martin_1997}, and/or heating of the ISM through interactions with the hot ICM or shocks \citep{poggianti_2019a,campitiello_2021}. Although such enhancements have been observed in higher-redshift jellyfish galaxies \citep[$z\sim0.3$–$0.4$;][]{moretti_2022}, our work provides the first clear evidence of this behaviour at low-redshift. \citet{moretti_2022} were unable to distinguish between the DIG and hot-ICM interaction scenarios as the dominant origin of the elevated ratios, due to the lack of data required to do so.

\subsection{High Emission Line Velocities and Dispersions}

In tail regions, we observe minimal or no stellar emission, confirming the absence of a stellar population in these regions. In contrast, the tail emission line velocities and velocity dispersions are significantly elevated relative to the disk values, particularly along the edges of the tails. We see the same or more extreme velocity dispersion trends in the other emission lines studied in this paper, with the mean velocity dispersion across all lines in the tail being $1.8\pm1.1\times$ higher than the mean in the disk. Shock modelling predicts high emission line velocities, coupled with emission line velocity dispersions on the order of the shock velocity \citep[e.g.,][]{dopita_1995,allen_2008,rich_2011,sutherland_2017}, in agreement with our average measured emission line velocities of $\sim45$ and $\sim150~\mathrm{km/s}$ in the disk and tail, respectively, and corresponding average dispersions of $\sim85$ and $\sim125~\mathrm{km/s}$.

Our measured tail velocities ($\sim150~\mathrm{km/s}$) are weakly supersonic compared to the typical ISM sound speed of $\sim10$–$100~\mathrm{km/s}$ (assuming ideal gas and Milky Way ISM temperatures), consistent with low-Mach shocks seen in radio spectral indices. In a study of 25 jellyfish galaxies in the Coma cluster, with some overlap with our sample, \citet{roberts_2024b} find radio spectral indices indicative of low-Mach number shocks or long electron-acceleration timescales. They also find that the bulk gas velocities in the tails are on the order of several hundred kilometres per second, consistent with our emission-line measurements. We note that both measurements are line-of-sight/projected velocities, and are therefore lower limits on the true 3D velocities.

In a study of 7 jellyfish galaxies in GASP, \cite{radovich_2019} find similarly elevated emission line velocities, including one galaxy with a high [\ion{O}{1}]/H$\alpha$ ratio and [\ion{N}{2}]/H$\alpha$ and [\ion{S}{2}]/H$\alpha$ ratios consistent with star formation. This galaxy, JO175, is well-described by their shock model, and is the only galaxy in their sample that does not host a central AGN, in-line with the properties of our sample.

Taken together, our observations favour a shock-ionization scenario for the ionized gas in radio-identified jellyfish galaxy tails in the Coma cluster. While LINER-like emission and high [\ion{O}{2}]/H$\alpha$ ratios can each arise from varying mechanisms, the consistent presence of elevated velocity dispersions and the spatial coincidence of these features in the tails denote shocks as the most likely common driver. The full CLIFS sample will enable statistical constraints on the incidence and strength of RPS-driven shocks across the Coma cluster.

\section{Summary and Conclusions}
\label{sec:summary}

In this work, we analyze a preliminary sample of 5 jellyfish galaxies in the Coma cluster using WEAVE LIFU data from the CLIFS survey. Emission line properties were extracted using the MaNGA Data Analysis Pipeline, allowing us to compare disk and tail regions to investigate the physical processes responsible for ionized gas in the tails.

We first compared SFRs derived from H$\alpha$ and from 144~MHz radio continuum emission, finding agreement in disk regions but significantly elevated radio ``SFRs'' in tail regions. This, coupled with spectral index work \citep{roberts_2024b}, suggests that processes other than star formation alone are exciting the gas in the tails. Emission line diagnostic diagrams further support this interpretation: tail spaxels are typically classified as star-forming in the [\ion{N}{2}] and [\ion{S}{2}] BPT diagrams, but are classified as LINER emission in the [\ion{O}{1}] diagram, consistent with excitation by shocks driven by ram pressure stripping. This scenario is further supported by enhanced [\ion{O}{2}]/H$\alpha$ ratios, as well as systematically elevated emission line velocities and velocity dispersions in the tails relative to the disks.

Once complete, the CLIFS survey will provide an unbiased sample of 29 jellyfish galaxies in the Coma cluster, selected without the requirement of detectable H$\alpha$ tail emission. The primary goals of the CLIFS survey are to: (1) constrain the fraction of LOFAR-selected radio tails that show co-spatial H$\alpha$ emission, and (2) spatially resolve both the star formation histories and the 144 MHz radio luminosity versus SFR relation across the star-forming disks. Using a sample of LOFAR jellyfish, including 16 in the Coma cluster, \citep{roberts_2022} found evidence for star formation on the leading halves of the disks. The complete CLIFS sample will enable us to extend this analysis to the entirety of the Coma cluster, and compare against CLIFS and MaNGA observations of undisturbed star-forming Coma cluster galaxies as a control.

\begin{acknowledgments}
We thank the anonymous referee for their useful and insightful comments. LMF and LCP thank the Natural Science and Engineering Research Council for funding. IDR acknowledges support from the Banting Fellowship Program. AI acknowledges funding from the European Research Council (ERC) under the European Union's Horizon 2020 research and innovation program (grant agreement No. 833824).

Based on observations made with the William Herschel Telescope operated on the island of La Palma by the Isaac Newton Group of Telescopes in the Spanish Observatorio del Roque de los Muchachos of the Instituto de Astrofísica de Canarias (WEAVE proposal(s) WS2023B2-001).

Funding for the WEAVE facility has been provided by UKRI STFC, the University of Oxford, NOVA, NWO, Instituto de Astrofísica de Canarias (IAC), the Isaac Newton Group partners (STFC, NWO, and Spain, led by the IAC), INAF, CNRS-INSU, the Observatoire de Paris, Région Île-de-France, CONACYT through INAOE, the Ministry of Education, Science and Sports of the Republic of Lithuania, Konkoly Observatory (CSFK), Max-Planck-Institut für Astronomie (MPIA Heidelberg), Lund University, the Leibniz Institute for Astrophysics Potsdam (AIP), the Swedish Research Council, the European Commission, and the University of Pennsylvania.  The WEAVE Survey Consortium consists of the ING, its three partners, represented by UKRI STFC, NWO, and the IAC, NOVA, INAF, GEPI, INAOE, Vilnius University, FTMC – Center for Physical Sciences and Technology (Vilnius), and individual WEAVE Participants. The WEAVE website can be found at \url{https://weave-project.atlassian.net/wiki/display/WEAVE} and the full list of granting agencies and grants supporting WEAVE can be found at \url{https://weave-project.atlassian.net/wiki/display/WEAVE/WEAVE+Acknowledgements.}

The authors acknowledge the use of the Canadian Advanced Network for Astronomy Research (CANFAR) Science Platform operated by the Canadian Astronomy Data Centre (CADC) and the Digital Research Alliance of Canada (DRAC), with support from the National Research Council of Canada (NRC), the Canadian Space Agency (CSA), CANARIE, and the Canada Foundation for Innovation (CFI).
\end{acknowledgments}

This research has made use of the NASA/IPAC Extragalactic Database (NED), which is operated by the Jet Propulsion Laboratory, California Institute of Technology, under contract with the National Aeronautics and Space Administration.

\facilities{WEAVE, LOFAR}

\software{\texttt{astropy} \citep{astropy_2013,astropy_2018,astropy_2022}, the MaNGA DAP \citep{belfiore_2019,westfall_2019}, \texttt{imfit} \citep{erwin_2015}, \texttt{matplotlib} \citep{hunter_2007}, \texttt{numpy} \citep{harris_2020}, \texttt{pandas} \citep{mckinney_2010, reback_2020}, \texttt{photutils} \citep{bradley_2024}, \texttt{scipy} \citep{virtanen_2020}, \texttt{seaborn} \citep{waskom_2021}, and \texttt{topcat} \citep{taylor_2005}.}

\appendix
\section{BPT Diagrams}
\label{appendix}

\begin{figure*}
    \centering
    \includegraphics[width=\linewidth]{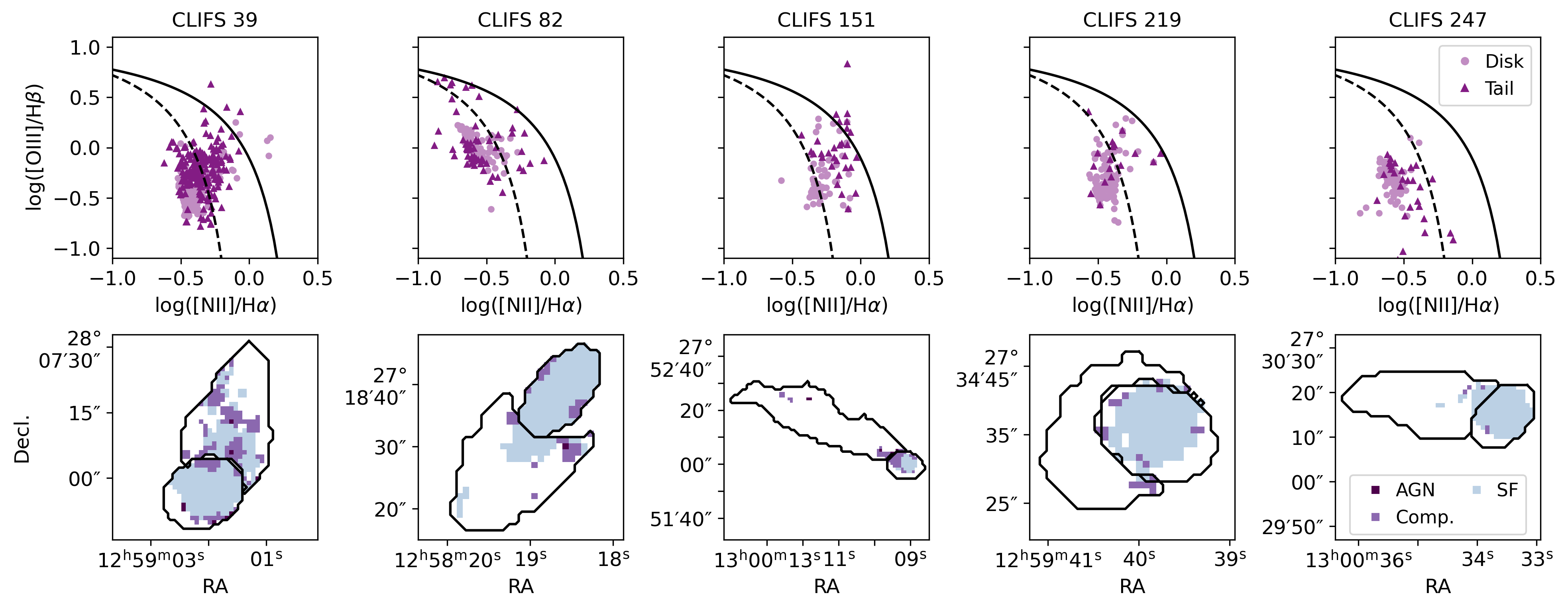}
    \caption{Top row: Resolved [\ion{N}{2}]/H$\alpha$ BPT diagram for each galaxy in the sample. Colours and cuts made are the same as in Figure \ref{figure_BPT}. Here, the solid and dashed lines are those defined by \cite{kewley_2001} and \cite{kauffmann_2003} respectively to distinguish between star formation (left), composite (centre), and AGN (right) as emission sources. Bottom row: Galaxy maps coloured according to the BPT classification above, with axes and contours as in Figure \ref{figure_BPT}. Star-forming, composite, and AGN pixels are coloured blue, light purple, or dark purple respectively.}
    \label{figure_BPTb}
\end{figure*}

\begin{figure*}
    \centering
    \includegraphics[width=\linewidth]{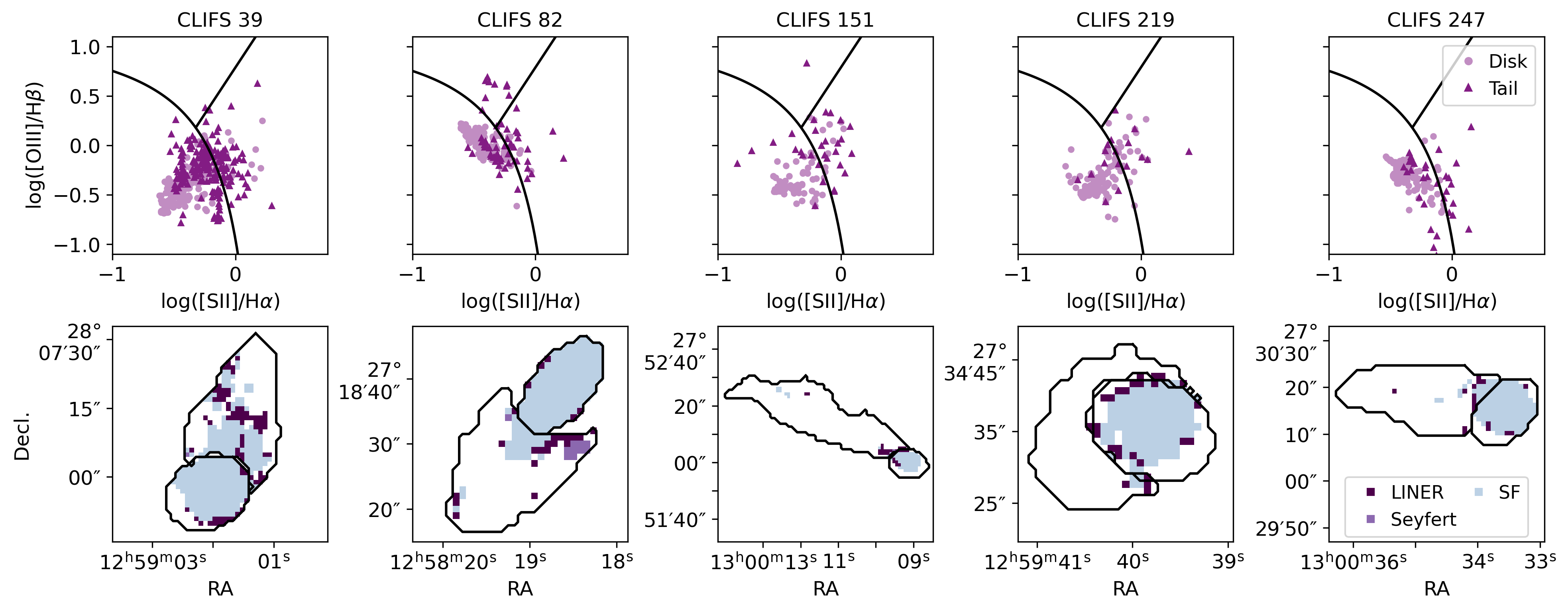}
    \caption{Top row: Resolved [\ion{S}{2}]/H$\alpha$ BPT diagram for each galaxy in the sample. Colours, cuts made, and classification boundaries are the same as in Figure \ref{figure_BPT}. Bottom row: Galaxy maps coloured according to the BPT classification above, as in figure \ref{figure_BPT}.}
    \label{figure_BPTc}
\end{figure*}

Here we provide the [\ion{N}{2}] (Figure \ref{figure_BPTb}) and [\ion{S}{2}] (Figure \ref{figure_BPTc}) BPT diagnostics for our sample.

\bibliography{biblio}{}
\bibliographystyle{aasjournalv7}

\end{document}